# Adaptation of XAI to Auto-tuning for Numerical Libraries

Shota Aoki
*Graduate School of Informatics*
Nagoya University
Aichi, Japan
aoki@hpc.itc.nagoya-u.ac.jp

Takahiro Katagiri
*Information Technology Center*
Nagoya University
Aichi, Japan
katagiri@cc.nagoya-u.ac.jp

Satoshi Ohshima
*Research Institute for Information Technology*
Kyushu University
Fukuoka, Japan
ohshima@cc.kyushu-u.ac.jp

Masatoshi Kawai
*Information Technology Center*
Nagoya University
Aichi, Japan
kawai@cc.nagoya-u.ac.jp

Toru Nagai
*Information Technology Center*
Nagoya University
Aichi, Japan
nagai@cc.nagoya-u.ac.jp

Tetsuya Hoshino
*Information Technology Center*
Nagoya University
Aichi, Japan
hosino@cc.nagoya-u.ac.jp

*Abstract*—Concerns have arisen regarding the unregulated utilization of artificial intelligence (AI) outputs, potentially leading to various societal issues. While humans routinely validate information, manually inspecting the vast volumes of AI-generated results is impractical. Therefore, automation and visualization are imperative. In this context, Explainable AI (XAI) technology is gaining prominence, aiming to streamline AI model development and alleviate the burden of explaining AI outputs to users. Simultaneously, software auto-tuning (AT) technology has emerged, aiming to reduce the man-hours required for performance tuning in numerical calculations. AT is a potent tool for cost reduction during parameter optimization and high-performance programming for numerical computing. The synergy between AT mechanisms and AI technology is noteworthy, with AI finding extensive applications in AT. However, applying AI to AT mechanisms introduces challenges in AI model explainability. This research focuses on XAI for AI models when integrated into two different processes for practical numerical computations: performance parameter tuning of accuracy-guaranteed numerical calculations and sparse iterative algorithm.

## I. Introduction

Concerns persist regarding the utilization of answers produced by artificial intelligence (AI) without sufficient verification, potentially leading to societal ramifications. Moreover, certain AI-generated outcomes are derived from publicly accessible web documents and other sources, lacking assurances of accuracy. Given the current AI landscape, human involvement in initial verification is necessary. However, manually reviewing the extensive output generated by AI models is impractical. Consequently, automating AI output verification processes and reducing associated human efforts are paramount. This has spurred extensive research and development endeavors aimed at minimizing the cost of constructing AI models.

Conversely, AI is revolutionizing software auto-tuning (AT) technology [1], aiming to streamline the labor-intensive task of performance tuning in numerical computation. This AT technology transcends mere parameter adjustments; it offers a holistic approach to performance optimization, encompassing the generation of high-performance code and addressing higher-level optimization objectives like compiler enhancements and algorithm selection. Our research in this domain specifically focuses on fine-tuning performance parameters.

Within this study, we will assess the explainability of AI-generated prediction times when implementing an AT function utilizing AI. We will utilize examples from performance parameter tuning in dense matrix libraries and sparse matrix iterative libraries to elucidate the explainability aspect.

This paper is organized as follows. Section II provides an overview of the tools used for predicting AI output. Section III details the case studies conducted in our research. Section IV presents the experimental results, and we conclude the paper in the final section.

## II. XAI Tools

### A. Overview

Numerous tools, commonly referred to as Explainable AI (XAI) tools, have been introduced to enhance the transparency and reliability of AI-generated results. Additionally, some of these XAI tools have been made openly accessible for easy utilization. Here, we provide an overview of some typical XAI tools.

### B. LIME

LIME (Local Interpretable Model-agnostic Explanations) [2] stands out as a noteworthy XAI tool suitable for a wide array of trained models. It operates as a local surrogate model, designed to elucidate the rationale behind individual predictions generated by black box models. Its primary function lies in presenting the justification for a specific classifier's prediction in a comprehensible manner for humans. By assessing the contribution of each feature to the classification process, LIME provides insights into the classifier's prediction outcomes. A distinctive feature of LIME is its versatility, as it can be applied seamlessly to any classifier, leveraging the classifier's prediction results. Essentially, LIME serves as a "local explanation tool," offering insights into individual cases.

*C. SHAP*

Aside from LIME, another widely adopted XAI tool is SHAP (SHapley Additive exPlanations) [3]. SHAP represents an application of the Shapley Value concept derived from cooperative game theory in the realm of machine learning. Consequently, the criteria employed for evaluation in SHAP have a strong theoretical foundation. It is worth noting that calculating the Shapley value, when approached rigorously, demands a significant computational effort. To address this challenge, research has focused on devising methods for approximate Shapley value calculation, which are effectively implemented in SHAP.

SHAP functions as a comprehensive explanatory tool, delving into overarching trends. It extends beyond elucidating individual predictions, offering a broader perspective on the behavior of machine learning models.

Within this research, our emphasis lies in the realm of global explanation using SHAP.

### III. CASE STUDY

*A. Aculeate Matrix-Matrix Multiplications*

*a) Ozaki Method*

In our research, we focus on a high-precision matrix-matrix library developed using Ozaki method [4]. Ozaki method involves the following processes, performed on input matrices while utilizing the IEEE754 standard floating-point data type:

1. Decompose matrix A and matrix B as illustrated in equation (1). Note that the number with the smaller value enclosed in parentheses corresponds to the higher-order bits:

$$A = A^{(1)} + A^{(2)} + A^{(3)} + \cdots + A^{(p)}$$
$$B = B^{(1)} + B^{(2)} + B^{(3)} + \cdots + B^{(q)} \quad (1)$$

2. Calculate the matrix product $A\,B$ as shown in equation (2).

$$A\,B = (A^{(1)} + A^{(2)} + A^{(3)} + \cdots + A^{(p)})(B^{(1)} + B^{(2)} + B^{(3)} + \cdots + B^{(q)})$$
$$= A^{(1)}B^{(1)} + A^{(2)}B^{(1)} + \cdots + A^{(p)}B^{(q)} \quad (2)$$

In this context, the decomposed matrix products as described in equation (2) are combined using a highly precise summation operation [4]. This meticulous approach ensures that a high level of calculation accuracy is maintained throughout the process. By developing a decomposition method in Process 1, it becomes feasible to render the matrix product in Process 2 an error-free transformation. Consequently, the transformation depicted in equation (1) is referred to as error-free transformation [4].

Our primary focus is on the computational characteristics of this error-free transformation in Process 1. In this process, the generation of a sparse matrix is contingent upon the degree of dispersion (range) observed in the element values' sizes within the input matrix. In this context, sparseness denotes the ratio of zero elements to the total element count within a dense matrix.

Due to these considerations, Ozaki method allows for the optimization of computations. Even when dealing with a dense matrix as an input, if the decomposition process through error-free transformation leads to a high degree of sparsity in the decomposed matrix, it becomes more efficient to perform calculations by converting to "sparse matrix-dense matrix product" and "sparse matrix-sparse matrix product" operations [5]. In essence, transitioning to these operations when implementing equation (1) helps reduce the overall computational load and execution time.

Currently, ongoing research is focused on proposing and evaluating the performance of a GPU implementation of Ozaki method [5], aiming to further enhance computational efficiency.

*b) Implementation of already developed*

In this study, we will utilize Ozaki's high-precision matrix multiplication library, *DHPMM_F*, specifically designed for GPU: High-precision Matrix Multiplication with Faithful Rounding [for GPU] (version 1.0) [5]. We aim to optimize its performance by considering the following 14 implementation choices as performance parameters to be fine-tuned. The details provided in parentheses pertain to the specific implementation details of the sparse matrix calculation method:

- Implementation on CPU:

    1. Implementation 1 (BLAS dgemm call (dgemm))

    2. Implementation 2 (Compressed Row Storage (CRS), Sparse Matrix-vector Multiplication (SpMV) - internal parallel)

    3. Implementation 3 (CRS, SpMV - external parallel)

    4. Implementation 4 (CRS, SpMV - multiple right-hand sides/internal parallel)

    5. Implementation 5 (CRS, SpMV - multiple right-hand sides, internal parallel blocking)

    6. Implementation 6 (Ellpack Format (ELL), SpMV - internal parallelism)

    7. Implementation 7 (ELL, SpMV - external parallelism)

8. Implementation 8 (ELL, SpMV - multiple right-hand sides, internal parallelism)
9. Implementation 9 (ELL, SpMV - multiple right-hand sides/internal parallelism blocking)
- Implementation on Graphics Processing Unit (GPU):
10. Implementation 10 (implementation using batched BLAS call, dense matrix operation)
11. Implementation 11 (dgemm, dense matrix operation)
12. Implementation 12 (CRS, SpMV)
13. Implementation 13 (ELL, SpMV)
14. Implementation 14 (CRS, SpMM)

In the context of the "blocking" approach, it is employed during SpMV implementation in sparse matrix-sparse matrix multiplication. However, it is executed as SpMV with multiple right-hand sides, where the number of multiple right-hand sides determines the block width [5]. This approach enables the efficient reuse of sparse matrix elements during SpMV for SpMV operations on a single right-hand side. As a result, this implementation method effectively mitigates cache misses when accessing data from the right-hand side vector, ultimately leading to improved performance.

In this case study, we will consider an AT mechanism using AI that automatically selects these 14 implementations.

B. PICCG Method

   a) Overview

Incomplete Cholesky Decomposition (IC) is employed in the context of systems of linear equations, represented as $Ax = b$, where $A \in \mathbb{R}^{n \times n}$, and $x, b \in \mathbb{R}^n$ with the matrix $A$ being sparse and symmetric. IC serves as a preprocessing step for the Conjugate Gradient (CG) Method, which is an iterative technique for solving systems of simultaneous linear equations. When the CG method is coupled with IC decomposition as a preprocessing step, it is referred to as the PICCG method.

Consider the application of IC for the decomposition of matrix A, as illustrated in equation (3) below:

$$A = U^t D\, U + R \ (U, D \in \mathbb{R}^{n \times n}) \quad (3)$$

In this context, we define the following matrices:
- $U$: An upper triangular matrix.
- $D$: A diagonal matrix.
- $R$: A matrix representing the difference between $A$ and $U^t D\, U$ after the IC decomposition process.

During the IC decomposition, elements in $A$ with initial values of 0 may transform into non-zero elements in the decomposition matrix $U$. This is referred to as "fill-in."

The fundamental IC decomposition approach is designed to minimize or reject fill-ins altogether, treating all such elements as zeros. This strategy aims to maintain a low count of non-zero elements in the matrix post-decomposition, thereby conserving memory and computational resources. However, it is important that if the disparity between $A$ and $U^t D\, U$ becomes significant, this approach may not yield optimal results.

   b) IC decomposition preprocessing and algorithm with threshold

Now, we will elucidate the IC decomposition method with the introduction of a threshold. In the matrix decomposition process described in equation (3), the non-zero elements within the decomposition matrix $U$ exhibit a fill-in level. This fill-in level signifies the extent to which non-zero elements are allowed to proliferate. Typically, the fill-in level is defined either on a per-row or per-diagonal element basis, considering the specific characteristics of the matrix or the problem.

In our research, we adopt the implementation by Kawai [6]. In this implementation, we configure two key parameters: the maximum fill-in level ($m$) and the threshold ($t$).

At the maximum fill-in level, elements that produce fill-ins below the specified level, with values smaller than the threshold, are treated as zeros, while fill-ins equal to or exceeding the threshold are permitted. This approach is devised to achieve convergence rates comparable to or better than those obtained with traditional IC decomposition preprocessing, while simultaneously minimizing the number of non-zero elements in the resulting matrices. This, in turn, enhances computational speed and maintains efficient memory usage. In essence, increasing the maximum fill-in level generally leads to fewer iterations, but as the matrix density and computational load increase, there exists a trade-off between iteration reduction and increased computation. These parameters can be fine-tuned for optimal performance. Empirically, a maximum fill-in level of 2 has proven optimal for the problems we are addressing. However, the selection of these parameters is contingent upon the specific nature of the problem under consideration.

An overview of the algorithm is depicted in equation (4), where $a_{i,j}$ denotes the element at row $i$ and column $j$ in matrix $A$, $d_{i,i}$ represents the diagonal element of $D$, $u_{i,j}$ signifies the element at row $i$ and column $j$ in matrix $U$, $f_{i,j}$ and $u_{i,j}$ represent the fill-in level, $t$ is the threshold value considered as zero, and $m$ denotes the maximum fill-in level.

In this case study, we will consider an AT mechanism using AI that automatically selects the $t$ (the threshold value considered as zero), and the $m$ (the maximum fill-in level).

$$d_{i,j} = a_{i,j} - \Sigma_{k=1}^{i-1} u_{i,k}\, d_{i,i}\, u_{k,j}$$
$$f_{i,j} = \{\, 0 \ (if\ a_{i,j} \neq 0)\, ,\, f_{i,k} + f_{k,i} + 1 \ (else)\,\}$$

$$u_{i,j} = \{ d_{i,j}^{-1}(a_{i,j} - \Sigma_{k=1}^{i-1} u_{i,k} d_{i,i} u_{k,j}) \ (if \ f_{i,j} \leq m \ \wedge |u_{i,j}| \geq t),$$
$$0(else) \} \quad (4)$$

## IV. EXPERIMENTAL RESULT

### A. Aculeate Matrix-Matrix Multiplications

*a) Experimental environment*

To acquire training data for our AI model, we use the computational power of the supercomputer "Flow" Type II subsystem [5], installed at Information Technology Center, Nagoya University. SHAP version 0.39.0 was used.

*b) Input data and AI model (optimal Implementation Selection)*

For our input data, which comprises matrices A and B, we followed two distinct generation methods:
1. Matrix elements were randomly generated within the range of 0 to 1. Elements with a certain level of sparsity had values inserted using the formula pow (10, rand() % $\Phi$), where $\Phi$ represents the upper limit, capped at 30.
2. A unit matrix and another matrix, whose elements were randomly generated within the range of 0 to 1, were used for elements with specific sparsity levels, ranging from 90% to 98%.

Our AI model, a classifier, is based on the random forest algorithm implemented in scikit-learn (version 0.24.1). Its primary task is to predict the fastest among the 14 specified implementations detailed in Section II. The matrices were of varying sizes, including 1000, 1500, 2000, 2500, and 3000 for both matrices A and B. In addition, matrix B had a size of 4000 in some cases. The training dataset comprised 68 samples, while the test dataset consisted of 13 samples.

The target variable for our classifiers the optimal implementation method among the 14 options. To facilitate this prediction, we provided the classifier with seven explanatory variables:
1. Matrix size.
2. Sparsity of matrix A.
3. Maximum value within matrix A.
4. Minimum value among the elements of matrix A.
5. Sparse decomposition counts of matrix A.
6. Count of dense decomposition matrices of matrix A.
7. Count of decompositions of matrix B.

The accuracy of our trained model varied, ranging from approximately 46% to about 92%, contingent on the data collection methodology employed.

*c) Case 1 : Optimal Implementation Selection*

Figure 1 shows the explanation results using SHAP.

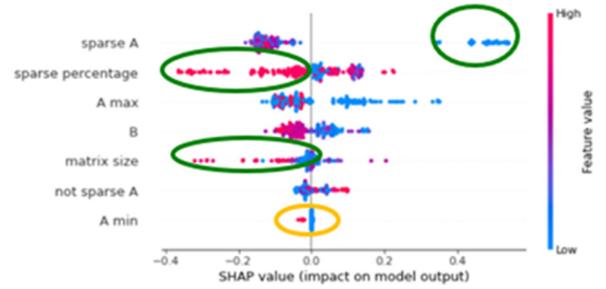
(a) CPU: Implementation 1 (dgemm)

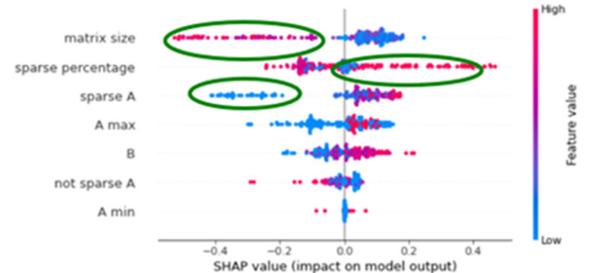
(b) CPU: Implementation 4 (CRS, SpMV - multiple right-hand sides/internal parallel)

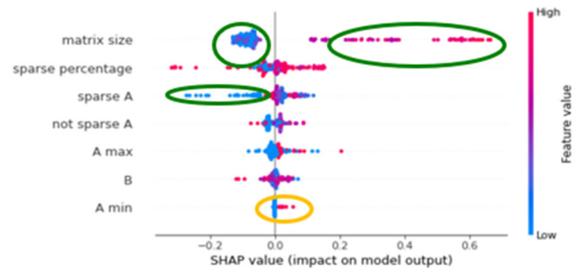
(c) GPU: Implementation 14 (CRS, SpMM)

Fig. 1. SHAP output for Implementation 1, 4, and 14.

In the SHAP output in Figure 1, we can derive meaningful insights by examining the clustering of explanatory variables where values are plotted closely together or where points with the same explanatory variable values (indicated by the same color) align in a pattern. The follows are summary of the interpretations from Figure 1:

1. **Sparse A**: The low sparsity of matrix A, indicating that it is close to a dense matrix, appears to strongly influence the choice of (a) implementation 1 (dgemm). This interpretation is coherent with the nature of implementation 1, as it primarily deals with dense matrix operations. On the other hand, the high sparsity of matrix A, indicating that it is close to a sparse matrix, appears to strongly influence the choice of (b) Implementation 4 (CRS, SpMV).

2. **Sparse percentage**: A high value for the sparsity percentage appears to be a factor against selecting (a) implementation 1 (dgemm). This interpretation aligns with the expectations for dense matrix operations, where high sparsity might not be the most efficient choice, while low sparsity might not be the most efficient choice for (b) CPU: Implementation 4 (CRS, SpMV).
3. **Matrix size**: The high value of the matrix size seems to deter the selection of (a) implementation 1 (dgemm). This interpretation is reasonable because this is dense matrix computation in CPU. On the other hand, the high value of matrix size seems to be critical for the selection (c) Implementation 14 (CRS, SpMM) in GPU. This is because, it can hide overhead of calling of GPU.
4. **Minimum value of elements of $A$**: This variable does not appear to significantly influence the choice of all implementations (a)-(c). This result suggests that, in this dataset, the elements of the matrix frequently become 0 when subjected to sparsification using error-free transformation. Consequently, the minimum value of the elements of $A$ often ends up being 0. As such, it does not serve as an effective explanatory variable for making predictions.

The point 4 underscores the importance of the test data. It highlights that sometimes; meaningful insights can only be revealed using XAI tools. AT library designers might not have prior knowledge of these subtleties, making the XAI tool an asset for uncovering hidden factors that impact model predictions. Thus, this example showcases the effectiveness of XAI tools in enhancing our understanding of complex models.

*d) Case 2: Setting the best block length*

Figure 2 illustrates the outcomes of explaining the influences of block width on CPU performance using SHAP. In Figure 2, the matrix size is fixed at 1500, while all other factors remain constant, except for sparsity, block width, and the number of decompositions of matrix $B$. Additionally, we employed a training dataset comprising 384 data points. The maximum relative error observed in our predictions was approximately 1.2%.

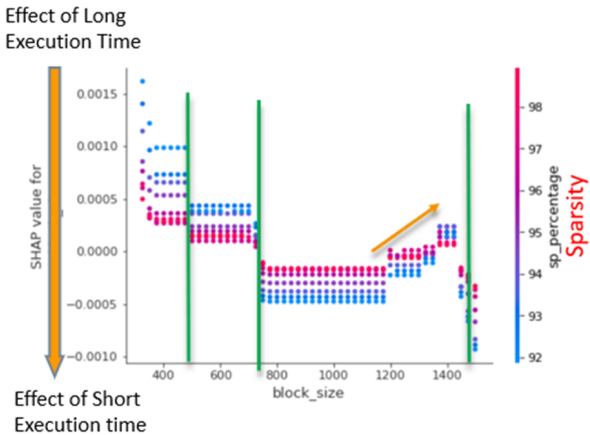

Fig. 2. SHAP output for blocking evaluation.

Figure 2 provides several valuable insights as follows:

1. When the block width is set to 700 or lower: A higher sparsity level tends to improve execution speed, while a lower sparsity level (indicating proximity to a dense matrix) leads to slower execution. This interpretation aligns with expectations because if the matrix is close to dense, it is likely that the data cannot fit efficiently into the cache memory given the specified block width.

2. When the block width exceeds 700: Smaller sparsity levels, closer to a dense matrix, result in faster execution speeds. This is a reasonable interpretation, as matrices resembling dense matrices benefit more from cache blocking techniques.

3. Within the block width range of 1200 to 1400: The block width appears to have a diminishing impact on speeding up operations. This could be due to the data size exceeding the cache capacity, rendering cache optimization less effective. This interpretation is sensible in such cases.

4. When the block width exceeds 1400: A larger block width contributes to improved execution speed. This interpretation holds true, especially when all the data fits comfortably within the cache, given the matrix size of 1500. However, the actual impact is typically influenced by hierarchical memory structures, and a detailed performance analysis is required to discern the specific reasons behind the speedup.

From the above observations, it becomes evident that SHAP provides meaningful explanations, particularly for users who may lack in-depth knowledge of cache blocking. The insight that the block width becomes a significant explanatory variable when it ranges from greater than 1200 to less than 1400 is particularly valuable. It underscores the effectiveness of SHAP in enhancing the understanding of complex performance factors.

*B. PICCG Method*

*a) Machine environment*

To gather AI training data, we leveraged the computational resources of the "Flow" Type I subsystem [8] at Information Technology Center, Nagoya University. For machine learning tasks, we use the GPU processing capabilities of the "Flow" Type II subsystem [7].

Our software stack included Python version 3.6.13, TensorFlow version 2.4.1, and SHAP version 0.39.0, which were used for conducting our machine learning experiments.

*b) Model to predict execution time of PICCG method*

For our machine learning model built using TensorFlow, we configured the following input and output:

- Input:
  1. Feature image of the coefficient matrix: These features were generated using the methodology proposed by Yamada et al. [9].

2. Maximum fill-in level and threshold: These parameters are crucial inputs for our model.

- Output:

  Computation time of the ICCG method with threshold: This is the target variable we aim to predict using our model.

We designed a Convolutional Neural Network (CNN) with the following architecture:

1. Convolutional Layers.
2. Pooling Layers: 2 layers.
3. Fully Connected Layers: 3 layers.

For our learning settings:
1. Number of Epochs: 200
2. Batch Size: 256
3. Activation Function: Rectified Linear Unit (ReLU)
4. Optimization Method: Adam's Method

Figure 3 provides an overview of the model's architecture and structure.

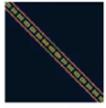

Fig. 3. Model architecture and structure.

*c) Creating a dataset*

The target application under consideration is *P3D* [10], which is designed for solving the steady-state heat conduction equation within a three-dimensional region. This region is discretized using a difference grid, and the application is based on the finite volume method [10].

Within this problem space, the distribution of thermal conductivity $\lambda$ is a key parameter, where $\lambda_2$ is constrained to be less than or equal to $\lambda_1$. A unique aspect of this problem is the introduction of a layer with thermal conductivity $\lambda_2$ into a layer characterized by thermal conductivity $\lambda_1$. As $\lambda_2$ decreases in value, the condition number of the resultant coefficient matrix increases. This phenomenon implies that as $\lambda_2$ becomes smaller, the problem becomes more challenging and approaches a malignant problem. Consequently, iterative solution methods face greater difficulty in solving it. Thus, the problem exhibits the remarkable characteristic that its nature can be controlled and manipulated by varying the value of $\lambda_2$.

The training dataset was meticulously generated by measuring the computation time using the "Flow" Type I subsystem under the following comprehensive conditions:

- Coefficient Matrix: Three different sizes were considered: 4096x4096, 32768x32768, and 262144x262144, totaling three types.

- Size Condition Number (Variation in Conductivity $\lambda_2$): There were 90 variations in the size condition number, reflecting changes in conductivity $\lambda_2$.

- Maximum Fill-in Level: Three maximum fill-in levels were considered: 0, 1, and 2.

- Threshold: Threshold values spanned from 0.001 to 0.02, with an interval of 0.001, resulting in 199 variations.

Overall, this exhaustive approach yielded a total of 41073 data points for each matrix size category. Additionally, a test dataset was created, comprising a total of 10269 data points for each matrix size.

The objective was to create a regression model for predicting the execution time of the ICCG method, considering the threshold, for each matrix size. This modeling effort allowed for a comprehensive analysis of the performance factors under varying conditions and enabled the development of predictive insights based on these intricate parameters.

*d) Explanation results using SHAP*

First, the validity of the generated AI model was further examined using SHAP explanations. Figure 4 provides insights into the explanation results for the problem sizes using SHAP.

Based on the interpretation derived from Figure 4, the following observations can be made:

1. When the threshold value falls below a certain threshold level, the execution time tends to decrease as the threshold value becomes smaller.

2. For the Fill-in level 2, the execution time tends to decrease when the threshold is set to 0.075 or lower in (b) 32768x32768.

These trends elucidated by SHAP explanations were not reliant on prior knowledge of the target algorithm. However, they align with the underlying algorithm of the ICCG method with IC preprocessing using a threshold, as well as the distribution pattern observed in the actual measured execution times of the test data. This alignment reinforces the validity of the AI model generated for this task. In essence, these findings affirm that the AI model is providing meaningful and valid insights, thereby enhancing the efficacy of the AT process.

On the other hand, we have found an unreasonable explanation by SHAP in viewpoint of PICCG algorithm. When the maximum fill-in level was either 0 or 1, there was no observable difference in the count of non-zero elements following the threshold IC decomposition within the dataset generated in this study. This observation can be attributed to the absence of elements with a fill-in level of 1. In simpler terms, even when adjusting the threshold to account for 0, no fill-in elements were eliminated. Consequently, the interpretation of the machine learning model's results relies on considering the threshold for 0, irrespective of whether the maximum fill-in level is 0 or 1.

To ensure an explanation that remains consistent and unaffected by the maximum fill-in level, we adopted the approach presented by SHAP. This approach asserts that as the maximum fill-in level increases, it becomes increasingly sensitive to the threshold considered for 0. In response, we treated the maximum fill-in level as qualitative data rather than quantitative data, generating a dummy variable. Subsequently, a new regression model was constructed using the preprocessed data, which included "One-Hot" encoding. Figure 5 shows the result by using the AI model generated with One-Hot encoding.

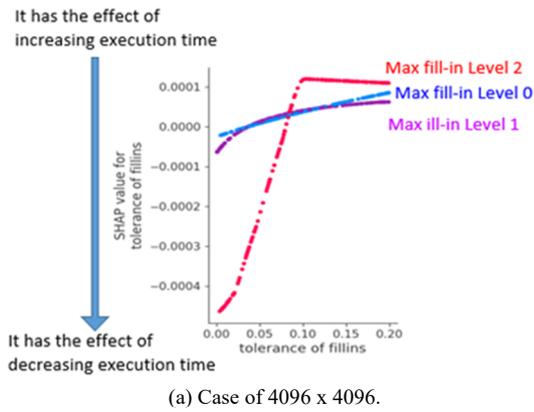

(a) Case of 4096 x 4096.

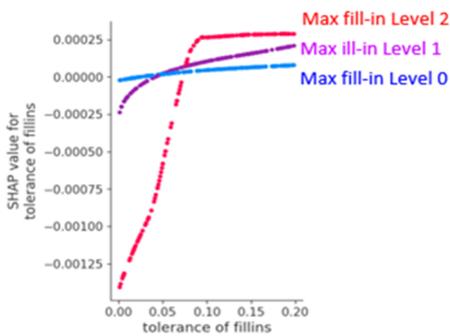

(b) Case of 32768 x 32768.

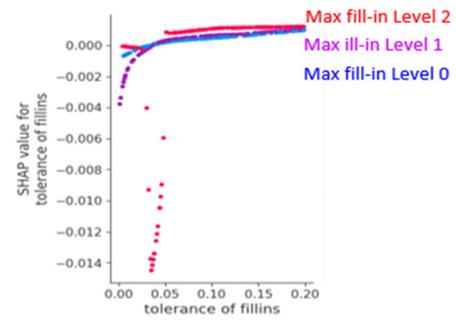

(c) Case of 262144 x 262144.

Fig. 4. SHAP output for Implementation 1, 4, and 14.

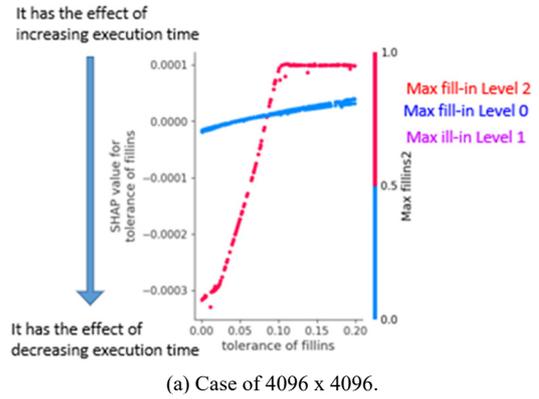

(a) Case of 4096 x 4096.

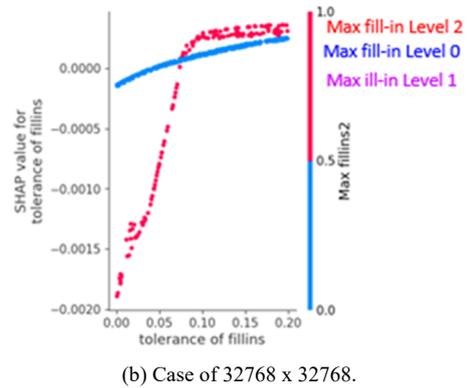

(b) Case of 32768 x 32768.

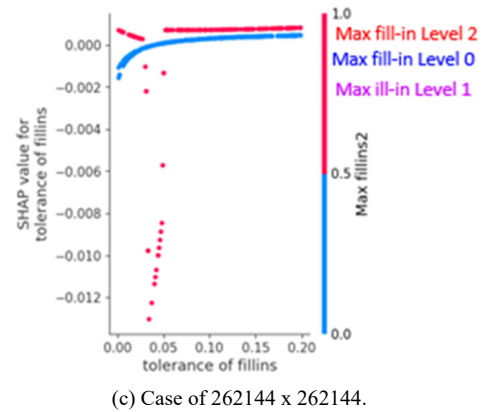

(c) Case of 262144 x 262144.

Fig. 5. SHAP output for performance parameters on PICCG method without dummy explanatory variables (One-Hot encoding).

The generated AI model in Figure 5 demonstrated its reasonability by achieving an average absolute error percentage of 2.73% (4096 x 4096), 2.64% (32768 x 32768), and 3.72% (262144 x 262144), which correspond to an error of less than 5%, when comparing the predicted execution time to the actual execution time. This achievement confirms that the model generation process was successful. Additionally, this signifies that within the scope of the test data, AI-assisted AT could be accomplished with an error margin of 5%.

In Figure 5, the SHAP illustration depicts a scenario where the influence impacting speed vanishes entirely at maximum fill-in levels of 0 and 1. This observation characterizes the specific attributes of the dataset in use, offering a plausible explanation. This instance exemplifies how examining SHAP explanations from an algorithmic standpoint can guide the enhancement of AI models.

Through this process of enhancing the explanatory variables within the AI model based on SHAP's insights, we achieved a noticeable improvement in prediction accuracy. Consequently, it can be concluded that XAI plays a crucial role in enhancing the capabilities of AI within numerical calculations in the field of auto-tuning.

## V. Conclusion

In this research, we utilized SHAP as an Explainable AI (XAI) tool to analyze the explainability of developing automated tuning (AT) functions leveraging AI for fine-tuning performance parameters, specifically within dense matrix computation and sparse matrix iterative solvers. Through a series of case studies, we derived findings and conclusions that offer valuable insights into SHAP's application in this domain.

The verification results yielded significant insights into the capabilities of XAI tools. Notably, SHAP explanations, in certain cases, provided well-founded and reasonable justifications for AT implementation selections. Moreover, our investigations revealed instances where these explanations effectively illuminated the optimal range of block widths for AT, enabling informed decisions regarding execution time utilizing block width as a crucial performance parameter.

Furthermore, XAI offered a meaningful interpretation by considering dataset characteristics from a numerical calculation algorithm perspective in the example illustrating performance parameter optimization for solving a sparse matrix iterative algorithm, despite lacking knowledge of the algorithm's specifics. These findings suggest the potential for XAI to provide valuable insights to developers of numerical computation libraries for debugging and performance enhancement purposes.

In future work, we will entail additional case analyses to demonstrate the effectiveness of XAI tools in the realm of AT functions for numerical calculations [11]. Additionally, considering the pressing concerns of AT for power consumption [12] and mixed-precision computation [13], we aim to explore the expansion of AT into these domains. We also plan to implement a mechanism for the automatic addition or reduction of effective explanatory variables, guided by insights gleaned from XAI tool explanations. This mechanism is poised to significantly enhance AT function performance, thereby advancing our capabilities in numerical computation processing.

ACKNOWLEDGMENT

This work was supported by JSPS KAKENHI Grant Number JP19H05662 and 23K11126, and by "Joint Usage/Research Center for Interdisciplinary Large-scale Information Infrastructures (JHPCN)" in Japan (Project ID: jh230005 and jh230053).